\documentclass[12pt]{iopart} 
 
\usepackage{epsfig} 
\usepackage{citesort}
      
\def\lsim{\lower0.6ex\vbox{\hbox{$ \buildrel{\textstyle <}\over{\sim}\ $}}}   
\def\gsim{\lower0.6ex\vbox{\hbox{$ \buildrel{\textstyle >}\over{\sim}\ $}}}      

\def\Msun{{\rm M}_{\odot}}
\def\Ni{^{56}{\rm Ni}}
\def\Co{^{56}{\rm Co}}
\def\Fe{^{56}{\rm Fe}}

\begin{document}   

\title[The Concordance Cosmic Star Formation Rate]
{The Concordance Cosmic Star Formation Rate: Implications from and 
for the Supernova Neutrino and Gamma Ray Backgrounds}     
   
\author{Louis E. Strigari$^{1}$, John F. Beacom$^{1,2}$, Terry P. Walker$^{1,2}$, 
and Pengjie Zhang$^{3}$}   
 
 \address{$^1$Department of Physics, The Ohio State University,    
Columbus, OH 43210}   
\address{$^2$Department of Astronomy, The Ohio State University,   
   Columbus, OH 43210}   
\address{$^3$NASA/Fermilab Astrophysics Center,
Fermi National Accelerator Laboratory, Batavia, IL 60510-0500}   
   
\ead{strigari@mps.ohio-state.edu, beacom@mps.ohio-state.edu,  
twalker@mps.ohio-state.edu, zhangpj@fnal.gov}   
  
\date{\today} 
   
\begin{abstract}       
  We constrain the Cosmic Star Formation Rate (CSFR) by requiring that
  massive stars produce the observed UV, optical, and IR light while
  at the same time not overproduce the Diffuse Supernova Neutrino
  Background as bounded by Super-Kamiokande.  With the massive star
  component so constrained we then show that a reasonable choice of
  stellar Initial Mass Function and other parameters results in SNIa
  rates and iron yields in good agreement with data.  In this
  way we define a `concordance' CSFR that predicts the optical SNII
  rate and the SNIa contribution to the MeV Cosmic Gamma-Ray
  Background.  The CSFR constrained to reproduce these and other
  proxies of intermediate and massive star formation is more clearly
  delineated than if it were measured by any one technique and has the
  following testable consequences: (1) SNIa contribute only a small
  fraction of the MeV Cosmic Gamma-Ray Background, (2) massive star
  core-collapse is nearly always accompanied by a successful optical
  SNII, and (3) the Diffuse Supernova Neutrino Background is
  tantalizingly close to detectability.

\end{abstract}   
   
\maketitle 
\section{Introduction} 

The Cosmic Star Formation Rate (CSFR), which encodes the history of
stellar birth, life, and death, is an essential probe of the evolution
of galaxies and the Universe. In recent years, the CSFR has been
probed out to redshifts $z \sim 6$, the results all indicating that
the CSFR was nearly an order of magnitude greater in the past than
today
\cite{Gallego95,Lilly96,Steidel96,Connolly97,Cowie99,Sullivan00,Scott02,Chapman03,Ouchi03,Stanway03,Heavens:2004sr}.
The CSFR is often measured by the total light output from young,
bright stars, though this can be significantly complicated by dust
obscuration. One can also probe the CSFR with the by-products of
stellar death: e.g., the neutrinos from core-collapse Type II
supernovae (SNII) for massive stars, and gamma rays from Type Ia
supernovae (SNIa) for some stars of intermediate mass.

Each SNII event creates a prompt burst of $10^{58}$ neutrinos, a
result which was confirmed by the detection of the neutrino flux from
SN 1987A in the Large Magellanic Cloud
\cite{Hirata:1987hu,Bionta:1987qt}. Though the neutrino flux from an
individual SNII beyond the local group of galaxies is below the
threshold of current detectors \cite{Ando:2005ka}, the cumulative
emission from all past SNII results in a {\it Diffuse Supernova
  Neutrino Background} (DSNB), which is on the verge of detectability.
Though the search for the DSNB flux is hindered by atmospheric
neutrino backgrounds, Super-Kamiokande has established an upper limit
on the electron anti-neutrino flux above 19.3 MeV of $1.2 \, {\rm
  cm}^{-2} \, {\rm s}^{-1}$ \cite{Malek:2002ns}, which nearly agrees
with recent theoretical predictions
\cite{Ando:2003tt,Fukugita:2003tt,Strigari:2003ig}.  The DSNB flux
probes the history of the SNII rate, about which little is directly
known, both locally and at high redshifts. Therefore, observations of
the high mass CSFR, corresponding to the range of SNII progenitors,
are generally used in predictions for the DSNB flux. The
Super-Kamiokande flux limit is in fact strict enough to rule out
scenarios for the CSFR, and provides a strong limit on the dust
corrections applied to UV light measurements.

By combining the DSNB flux limit with determinations of the CSFR from
light surveys and other data, we construct a more restrictive
concordance model for the CSFR.  More specifically, the DSNB sets an
upper limit on the cosmic SNII rate, and therefore the CSFR above the
core-collapse threshold mass of $8 \, \Msun$ \cite{Heger:2002by}.
Throughout this paper, we follow the convention of referring to
massive star core-collapse generically as SNII, including the less
dominant core-collapse SNIb and SNIc in this definition.  We assume
these stars end their lives in a prompt burst of neutrinos. Generally,
massive star core-collapse need not produce visible light, if the core
bounce shock fails to eject the stellar envelope and a black hole is
formed \cite{Heger:2002by,Beacom:2000ng,Beacom:2000qy}. When the
supernova is successful, we refer to it as an optical SNII. The
massive stars which emit UV light in their lifetimes also contribute
to the cosmic far infrared background (FIRB) flux, which is believed
to arise from the absorption and re-emission of UV light
\cite{Hauser:2001xs,Wright:2003tp}.  Measurement of the slope of the
CSFR from UV and IR light, combined with a normalization from the DSNB
flux, provides a tight constraint on the CSFR for M $> 8 \, \Msun$ and
$z \, \lsim \, 1$.

We determine the intermediate-mass CSFR from the high-mass CSFR using
the stellar {\it Initial Mass Function} (IMF), which is observed to
have a universal slope above $\sim 0.5 \, \Msun$. The
intermediate-mass ($\sim 3-8 \, \Msun$) CSFR has implications for the
rate of SNIa, though knowledge of the IMF and high mass CSFR does not
precisely fix the SNIa rate.  The lack of a well-defined progenitor
model introduces some ambiguity in the time delay between progenitor
formation and explosion, and the efficiency for creating SNIa from a
main sequence population.  However, these parameters can be fixed
empirically by requiring a progenitor model to reproduce the observed
evolution of the SNIa rate and the contribution to the mean iron
abundance in the universe.  We examine a range of allowed values for
these parameters, and use these to examine the evolution of the SNIa
rate.

Each SNIa event is expected to produce $\sim 10^{55}$ gamma rays, and
thus the SNIa rate can be tested by the SNIa contribution to the MeV
{\it Cosmic Gamma-Ray Background} (CGB). From the hard x-rays to high
energy gamma rays, the CGB spectrum is observed to have a steep
decline, and is typically explained as the sum of three components.
Unresolved active galaxies are believed to be the dominant sources
below $\sim 1$ MeV (Seyferts) and above $\sim 10$ MeV (Blazars). These
hypotheses are strengthened by the fact that observations of resolved
Seyferts and Blazars, combined with estimates of their number density,
can reasonably reproduce the spectrum normalization and shape in the
respective energy ranges \cite{Sreekumar:1997yg}.  In the range $1-3$
MeV, it has generally been assumed that SNIa account for the measured
CGB
\cite{clayton91,Watanabe:1998ds,Ruiz-Lapuente:2000zx,Zhang:2004tj}.
In this paper we show that the derived limits on the SNIa rate implies
that SNIa cannot be the principal source of the MeV CGB.
 
The paper is organized as follows.  In Section \ref{sec:CSFR} we
review current observations for the CSFR. In Section \ref{sec:dsnb} we
discuss the relation between the SNII rate and the DSNB. In Section
\ref{sec:SNIa}, we estimate the evolution of the supernova rates, and
in Section \ref{sec:indirect} we discuss estimates of the iron
abundance and the far infrared background. In Section \ref{sec:CGB} we
use the SNIa rate to estimate the SNIa contribution to the CGB, and in
Section \ref{sec:conclusions} we discuss some implications for the
DSNB and the CGB. 

\section{The Concordance Cosmic Star Formation Rate} 
\label{sec:CSFR}

Determining the CSFR at a particular redshift typically involves
either measuring a UV continuum or H$\alpha$ emission for a particular
galaxy, and then multiplying by the space density of galaxies at that
redshift \cite{Tresse:2001eb}.  These results depend most strongly on
stars of mass $\gsim 3 \, \Msun$, and require extrapolation of the IMF
to lower masses to obtain the CSFR for all masses.  The corrections
for UV extinction by dust are large, and at least until recently,
rather uncertain.

Complementing these methods are recent results from the 2-degree Field
survey (2dF) and Sloan Digital Sky Survey (SDSS), which measure the
ensemble spectrum of present day galaxies, thus determining the mix of
old and young stellar populations and fitting CSFR models to this
spectrum \cite{Glazebrook:2003xu,Baldry}.  These results are sensitive
to the entire range of stellar masses, and thus have a different (and
ideally less strong) dependence on the assumed dust extinction
corrections.  Given the limitations of existing data, in the analysis
of the 2dF and SDSS cosmic optical spectrum, the CSFR was parametrized
as a broken power law:
\begin{eqnarray}
{\rm R}_{{\rm SF}}  (z) & =  & {\rm R}_{{\rm SF}}(0) (1+z)^\beta \quad   
  {\rm for} \quad  z <
       z_{p}  \nonumber \\  & = & {\rm R}_{{\rm SF}}(z_{p}) \,  
(1+z)^\alpha  \quad  {\rm for}
       \quad z > z_{p}  ,
\label{eq:powerlaw}
\end{eqnarray}
where $\beta \sim 2$, $z_{p} \sim 1-2$ is the transition redshift, and
$\alpha \sim 0$.

\begin{figure}
\epsfxsize=5in
\begin{center}
\leavevmode
\epsffile{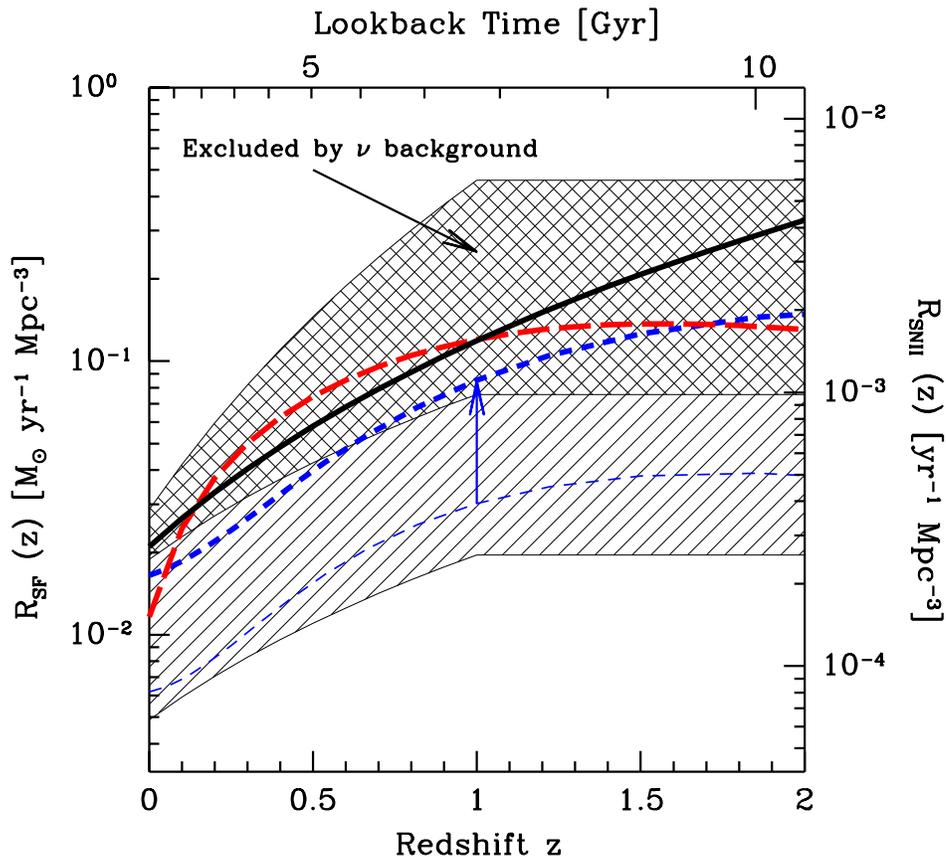}
\end{center}
\caption{
  Observational results for the Cosmic Star Formation Rate, with the
  conversion to SNII rate on the right axis using Equation
  (\ref{eq:SNIIRate}).  The entire cross-hatched plus shaded region is
  consistent with the results of the 2dF and SDSS cosmic optical
  spectrum \cite{Glazebrook:2003xu,Baldry}.  The upper cross-hatched
  region is ruled out by the limit on the DSNB flux, while the lower
  shaded region is allowed.  Three recent (dust corrected by those
  authors) results are also shown: long-dashed red line (Dahlen et
  al.~\cite{Dahlen:2004km}), solid black line
  (GALEX~\cite{Schiminovich:2004km}), and short-dashed blue line (Cole
  et al.~\cite{Cole:2001}). In the latter case we also show their
  result before dust correction~\cite{Cole:2001}; dust corrections in
  the other cases are similar. The concordance region is driven by the
  proximity of these recent observations to the upper bound from the
  neutrino data, and therefore is concentrated at the upper edge of
  the lower band.  }
\label{fig:CSFRmodels}
\end{figure}

In Figure~\ref{fig:CSFRmodels}, we show representative fits to
observations of the CSFR.  The entire cross-hatched plus shaded region
depicts the allowed region from the 2dF and SDSS cosmic optical
spectrum.  For $z < 1$, the upper curve has $\beta = 4$, and the lower
curve has $\beta = 2$.  For both the upper and lower limits we use
$\alpha = 0$ beyond $z_p = 1$; both $\alpha$ and $z_p$ are quite
uncertain from this data alone.  As examples of very recent UV
measurements, we show lines corresponding to the results of Cole et
al.~\cite{Cole:2001}, Dahlen et
al.~\cite{Dahlen:2004km,Strolger:2004kk}, and
GALEX~\cite{Schiminovich:2004km}.  All three results are corrected for
dust extinction by the authors; after correction, these UV results are
consistent with H$\alpha$ results.  We show also the uncorrected
results of Cole et al.~\cite{Cole:2001} as an example, as the
corrections of the other authors are similar.

%%%%%%%%%%%%%%%%%%%%%%%%%%%%%%%%%%%%%%%%%%%%%%%%%%%%%%%%%%%%%%%%%%%%%%%% 
%%%%%%%%%%
%%%%%%%%%%%%%%%%%%%%%%% CSFR models Figure  
%%%%%%%%%%%%%%%%%%%%%%%%%%%%%%%%%%%%%
%%%%%%%%%%%%%

The limit on the neutrino flux from the core collapse of massive stars
provides an important upper bound on the CSFR.  We emphasize that this
is an {\it integral} constraint, and does not provide {\it
  differential} sensitivity to the CSFR at each redshift.
Nevertheless, it is a strong constraint.  Following the 2dF and SDSS
analysis, we adopt the above four-parameter form for the CSFR.  Since
there is little sensitivity beyond $z \gsim 1$, we conservatively fix
$z_p = 1$ and $\alpha = 0$ (in Figure \ref{fig:CSFRmodels} we show
redshifts up to $z =2$ for illustration; less than $5\%$ of the DSNB
flux in the observable Super-K energy range comes from $z>1$).
Further, the UV and H$\alpha$ results do constrain $\beta$ well, so we
fix $\beta = 2$ (choosing $\beta$ on the small side gives a more
conservative limit on the normalization).  What remains is ${\rm
  R}_{{\rm SF}}(0)$, and the DSNB limit on this quantity is completely
independent of the issues associated with dust extinction.

The interesting tension is between the recent UV data, which has in
recent analyses been rising in normalization, and the DSNB limit,
which does not allow any further increase.  In fact, as we discuss
below, these dust corrected models are marginally inconsistent with
the DSNB upper limit; however, we view this discrepancy as within the
uncertainties.  Though the 2dF and SDSS band includes a region below
the dust corrected models, it is disfavored by combining all data
sets.  The concordance region is concentrated around the DSNB upper
limit, consistent with the light output of massive stars.  Thus the
``concordance" CSFR we obtain is principally defined by the upper
limit on the normalization ${\rm R}_{{\rm SF}}(0) \simeq 2 \times
10^{-2} M_\odot$ yr$^{-1}$ Mpc$^{-3}$ (with the assumed redshift
dependence of the CSFR).  On the other hand, recent astronomical data
probably do not permit a normalization below ${\rm R}_{{\rm SF}}(0)
\simeq 1 \times 10^{-2} M_\odot$ yr$^{-1}$ Mpc$^{-3}$.  While the
existing data do not warrant a more sophisticated statistical
analysis, it is clear that these combined results are much more
constraining than those of any one technique alone.

We use the results from Figure~\ref{fig:CSFRmodels} to estimate the
supernova rates, placing a strong focus on $z \, \lsim \, 1$, which is
the most relevant range for the supernova neutrino and gamma ray
backgrounds.  That is, in later figures the shaded bands shown
correspond exactly to the lower shaded band in
Figure~\ref{fig:CSFRmodels}; the region shown with the cross-hatched
band in Figure~\ref{fig:CSFRmodels} is considered excluded by the DSNB
limit.  Furthermore, as noted by our approximate constraint on ${\rm
  R}_{{\rm SF}}(0)$, the upper part of each shaded region is
preferred.

\section{The SNII Rate and the Diffuse Supernova Neutrino Background} 
\label{sec:dsnb}

Though the CSFR is observationally well-studied, relatively little is
known about the optical SNII rate, even locally.  The high mass and
short lifetimes of the progenitor stars imply that the SNII rate is
directly proportional to the CSFR at the same epoch. The DSNB has been
the subject of many previous studies, most recently
in~\cite{Ando:2003tt,Fukugita:2003tt,Strigari:2003ig,Beacom:2003nk}
(for earlier predictions,
see~\cite{Bisnovatyi,Krauss,Dar,Woosley,Totsuka,Totani95,Totani96,Malaney,Hartmann1,Kaplinghat:1999xi,Ando},
and for a recent review, see \cite{Ando:2004hc}).  In this section, we
review current estimations for the DSNB flux, updating for new results
for the CSFR.  For more details on the flux calculation we refer to
Strigari et al.~\cite{Strigari:2003ig}.

The DSNB flux is a convolution of the SNII rate with the neutrino
emission spectrum.  We denote the supernova rate as a function of
redshift as ${\rm R}_{{\rm SNII}} (z)$ (in units of number of
supernovae per time per comoving volume), and the neutrino spectrum
from an individual SNII event as $dN/dE$. The number flux of neutrinos
from SNII events is $4 \pi \, n$ (using $c = 1$ here and elsewhere),
where $n$ is the comoving neutrino number density, and the
differential flux in units of number of neutrinos per time $t$, per
area $A$, per energy is
\begin{equation} 
\frac{dN}{dt dA dE} = \int^{z_{{\rm max}}}_{0} {\rm R}_{{\rm SNII}}(z) \frac{dN(E(1+z))}{dE}
(1+z) \left \vert \frac{dt}{dz} \right \vert dz.
\label{eq:diffuseflux}
\end{equation} 
Here $E (1+z)$ is the neutrino energy at emission, $E$ is the energy
at detection, and $z_{{\rm max}}$ is the redshift at which star
formation begins. Here we take $z_{{\rm max}} =5$, though our results
are insensitive to this choice. We use a flat $\Lambda$CDM cosmology,
with $\Omega_{{\rm M}} = 0.3$ and ${\rm H}_{0} = 70 \, {\rm km} \,
{\rm s}^{-1} \, {\rm Mpc}^{-1}$ \cite{Turner:2002}, giving
\begin{equation}
\left \vert \frac{dt}{dz} \right \vert = \frac{1}
{{\rm H}_{0} (1+z) \sqrt{ \Omega_{{\rm M}} (1+z)^{3} + \Omega_{\Lambda}}}. 
\label{eq:dtdz}
\end{equation}  

The DSNB flux thus depends primarily on the SNII neutrino spectrum and
the SNII rate as a function of redshift.  In a SNII event, all flavors
of neutrinos are created with thermal spectra. Each neutrino flavor is
characterized by a temperature which depends on the radius at which
the neutrino decouples from the matter in the interior of the
proto-neutron star (for recent results on flavor dependent neutrino
fluxes from SNII, see~\cite{Keil:2002in}). SNII produce an order of
magnitude fewer gamma rays than SNIa, and they are attenuated further
by the thick stellar envelope.

The Super-Kamiokande upper limit on the flux of DSNB electron
antineutrinos above $19.3$ MeV is $1.2 \, {\rm cm}^{-2} \, {\rm
  s}^{-1}$. This is based on the limit on the rate of positron-like
events caused by DSNB neutrinos undergoing $\bar{\nu}_e + p \rightarrow
e^{+} + n$ \cite{Malek:2002ns}. As discussed in the previous section,
we use this integral constraint on the CSFR to bound ${\rm R}_{{\rm
    SF}} (0)$.

\begin{figure}
\epsfxsize=4.5in
\begin{center}
\leavevmode
\epsffile{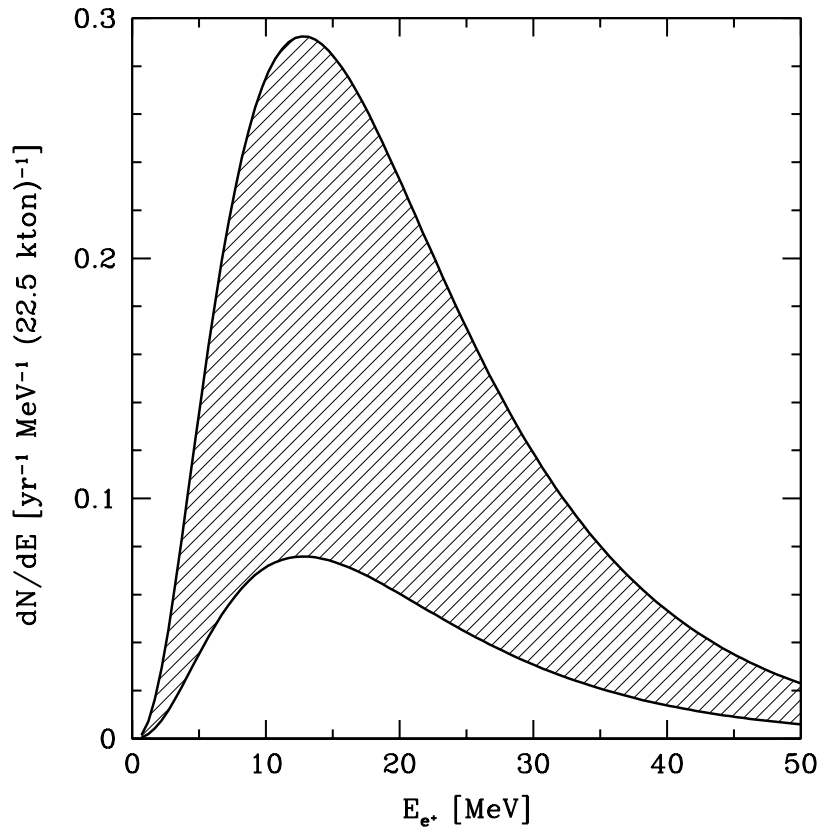} 
\end{center} 
\caption{The DSNB detection spectrum at Super-Kamiokande
  (i.e., weighted by the detection cross section for $\bar{\nu}_e +
  {\rm p} \rightarrow {\rm e}^{+} + {\rm n}$).  The shaded region
  shown corresponds to the similarly shaded region in
  Figure~\ref{fig:CSFRmodels}. The upper limit is determined by the
  Super-Kamiokande bound, and the lower limit by the 2dF and SDSS
  data. As noted, in the concordance model, the upper edge of the
  region is favored. There are significant detector backgrounds, which
  are discussed in~\cite{Strigari:2003ig,Ando:2003tt,Beacom:2003nk}.
}
\label{fig:Spectrum} 
\end{figure}

Converting the CSFR from Equation~\ref{eq:powerlaw} to the SNII rate
requires the IMF.  The IMF as measured in different environments is
consistent with a power law, $dn/dm \propto m^{-2.35}$, in the range
$m > 0.5 \, \Msun$.  The SNII rate as a function of redshift is then
\begin{equation}
{\rm R}_{{\rm SNII}} (z) = \int_{8}^{30} \frac{dn}{dm} \, {\rm R}_{{\rm SF}} (z) \, dm, 
\label{eq:SNIIRate}
\end{equation} 
where the units of mass are $\Msun$.  Here $dn/dm$ is the normalized
IMF, where for masses less than $0.5 \, \Msun$ we are using a slope
$m^{-1.50}$~\cite{Baldry:2003}.  We impose an upper cut-off to the
integral of 30 $\Msun$, though our results are insensitive to this
choice.  Thus $\sim 1\%$ of all stellar mass becomes core-collapse
supernovae.  In Figure~\ref{fig:Spectrum} we show the range of allowed
DSNB spectra, corresponding to the allowed region in Figure
\ref{fig:CSFRmodels}.
%%%%%%%%%%%%%%%%%%%%%%%%%%%%%%%%%%%%%%%%%%%%%%%%%%%%%%%%%%%%%%%%%%%%%%%%%%%%%%%%%%%
%%%%%%%%%%%%%%%%%%%%%%% Spectrum Figure %%%%%%%%%%%%%%%%%%%%%%%%%%%%%%%%%%%%%%%%%
%%%%%%%%%%%%%%%%%%%%%%%%%%%%%%%%%%%%%%%%%%%%%%%%%%%%%%%%%%%%%%%%%%%%%%%%%%%%%%%%%%%
%%%%%%%%%%%%%%%%%%%%%%%%%%%%%%%%%%%%%%%%%%%%%%%%%%%%%%%%%%%%%%%%%%%%%%%%%%%%%%%%%%%  

\section{The Type Ia Supernova Rate}
\label{sec:SNIa}

SNIa are believed to result from the thermonuclear detonation of a
white dwarf that has accreted the outer layer of its companion star to
the instability point.  In this model, the main sequence lifetime of
the progenitor star, combined with the accretion rate of the white
dwarf from the giant companion, imply a $\sim {\rm Gyr}$ delay time
between star formation and the SNIa event.  This time delay implies a
different relationship between the CSFR and SNIa rate compared to the
relation between the CSFR and the SNII rate. The result is that the
SNIa rate at a fixed redshift is an indirect tracer of the CSFR at
earlier epochs.
 
Additionally, SNIa result from a lower mass range of the stellar IMF,
approximately $3-8 \, \Msun$~\cite{Madau:1998}. The shape of the IMF
provides more stellar mass in this range relative to the range for
SNII progenitors, but SNIa are observed to be several times more rare
than their SNII counterparts, both locally and at moderate redshifts.
This result implies that the efficiency for creating SNIa is less than
that for SNII, which are expected to have an efficiency factor near
unity for stars greater than 8 $\Msun$ (see discussion below).
 
We parametrize the SNIa rate as
\begin{equation}
{\rm R}_{{\rm SNIa}} (t)  = \eta \int_{3}^{8} dm \, \frac{dn}{dm} \int^{t}_{0} dt_{Ia}
{\rm R}_{{\rm SF}} (t-t_{Ia}) g(t_{Ia}).
\label{eq:IaRate}
\end{equation}
Here $t_{Ia}$ is the delay time and $\eta \sim 1\%$ is the efficiency
of producing an SNIa event from stellar mass in the range $3-8 \,
\Msun$.  The delay time has been previously estimated by comparing the
observed CSFR to the SNIa rate as a function of
redshift~\cite{Strolger:2004kk,GalYam,Maoz:2003ee}, or from chemical
evolution models comparing the observed element enrichment as a
function of redshift~\cite{Yoshii:1996tt}.  These results indicate
mean time delays roughly in the range $1-3$ Gyr.  With the current
data on the SNIa rate, models with a constant time delay are not
distinguishable from more complicated models~\cite{Maoz:2003ee}.  Here
we consider a constant time delay of 3 Gyr, i.e. $g(t_{Ia})= \delta
(t_{Ia} - 3 \, {\rm Gyr}$), independent as well of the mass of the
progenitor star.

For this fixed time delay, we determine $\eta$ by comparing the
observed CSFR to the SNIa rate, and make the replacement in
Equation~\ref{eq:IaRate} that $\eta \, dn/dm \rightarrow f_{Ia}$. This
approach was taken by Ruiz-Lapuente et
al.~\cite{Ruiz-Lapuente:2000zx}, in comparing the SNIa rate to ${\rm
  R}_{{\rm SF}}$(z) at different redshifts. We additionally apply the
constraint on the CSFR above 8 $\Msun$ (or equivalently ${\rm R}_{{\rm
    SNII}}(z=0)$) from the DSNB flux.  Given the values of $\beta$ and
${\rm R_{{\rm SNII}}}(z=0)$ that maximize the DSNB flux, and a fixed
time delay of $ 3 \, {\rm Gyr}$, we determine that $f_{Ia} = (1/700 -
1/1000) \, \Msun^{-1}$ in order to remain consistent with the results
for the SNIa rate evolution. Motivated by establishing a robust upper
limit to the SNIa CGB flux below, we choose $(1/700) \, \Msun^{-1}$.

\begin{figure}
\epsfxsize=4.5in
\begin{center}
\leavevmode
\epsffile{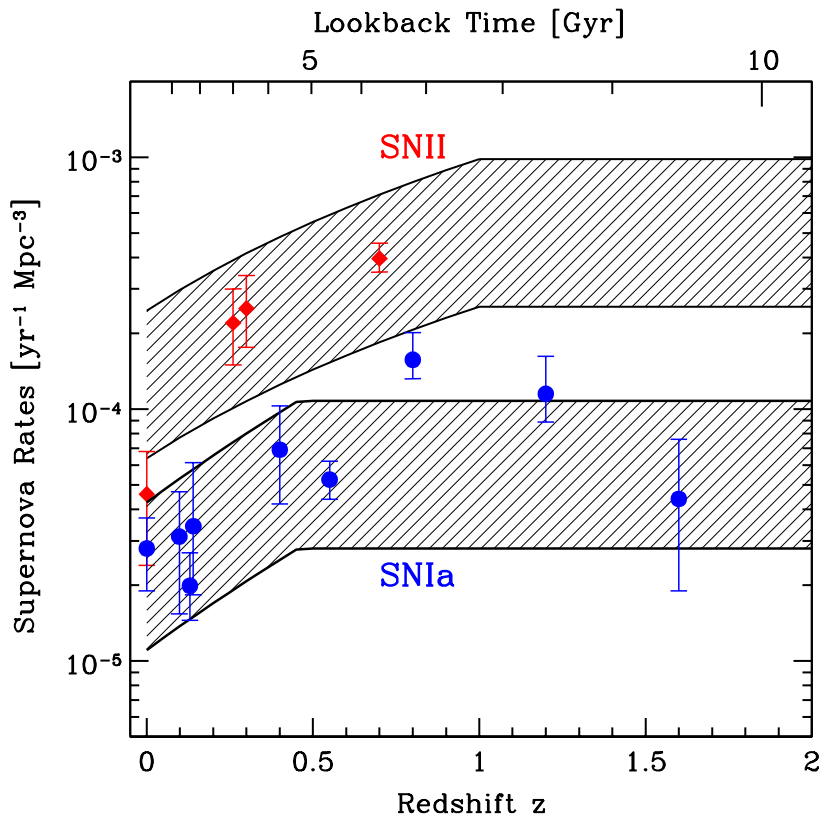} 
\end{center} 
\caption{
  The allowed bands for the supernova rates follow from the lower
  allowed band of Figure~\ref{fig:CSFRmodels}. The SNIa data is
  from~\cite{Dahlen:2004km} and~\cite{Blanc:2004ws}, and the optical
  SNII data is from \cite{Dahlen:2004km} and~\cite{Cappellaro:1999qy}.
  The error bars show statistical uncertainties. As discussed in Cappellaro
  et al.~\cite{Cappellaro:1999qy}, the $z = 0$ optical SNII measurement is
  not corrected for dust extinction, and represents a lower limit. 
  For the SNIa calculations, we use $f_{Ia} = (1/700) \, \Msun^{-1}$
  and a time delay of 3 Gyr.
\label{fig:SNrates} 
}
\end{figure}

In Figure~\ref{fig:SNrates}, we show our limits for the evolution of
the supernova rates in comparison to the observational results. The
allowed bands for the SNII and SNIa bands are correlated, in the
sense that increasing the SNIa contribution by increasing ${\rm R}_{{\rm SF}} (z=0)$
implies a similar increase in the SNII contribution, which would violate the DNSB flux limit. For
this figure we have used ${\rm R_{{\rm SF}}}(z=0) = 1.9 \times
10^{-2} \, \Msun \, {\rm yr}^{-1} \, {\rm Mpc}^{-3}$ and $\beta = 2$.

%%%%%%%%%%%%%%%%%%%%%%%%%%%%%%%%%%%%%%%%%%%%%%%%%%%%%%%%%%%%%%%%%%%%%%%%%%%%%%%%%%%
%%%%%%%%%%%%%%%%%%%%%%% SNIa Rates Figure %%%%%%%%%%%%%%%%%%%%%%%%%%%%%%%%%%%%%%%%%
%%%%%%%%%%%%%%%%%%%%%%%%%%%%%%%%%%%%%%%%%%%%%%%%%%%%%%%%%%%%%%%%%%%%%%%%%%%%%%%%%%%
%%%%%%%%%%%%%%%%%%%%%%%%%%%%%%%%%%%%%%%%%%%%%%%%%%%%%%%%%%%%%%%%%%%%%%%%%%%%%%%%%%%

Below we consider the CGB as an observable test of the SNIa rate.
Though changing the slope to $\beta = 4$ would require lowering ${\rm
  R_{{\rm SNII}}}(z=0)$ to remain consistent with the DSNB flux limit,
this does not appreciably change the SNIa CGB flux in the detectable
energy range $0.8-3.0$ MeV.  For a CSFR that increases or remains
constant beyond $z \sim 1$, the 3 Gyr time delay gives a maximal
contribution to the observable SNIa CGB flux. When considering a fixed
value for ${\rm R_{{\rm SNIa}}}(z=0)$, the time delay is degenerate
with the value of $f_{Ia}$, such that an increase in the time delay
requires a decrease in $f_{Ia}$. This results from the fact that
greater time delays sample the CSFR at higher redshifts.

\section{The Iron Abundance and Far IR Background} 
\label{sec:indirect}

\subsection{Iron Abundance}

Observations and models indicate that, on a per event basis, SNIa
produce an order of magnitude more iron than SNII~\cite{hamuy03}. With
the concordance CSFR, the integrated SNII number density is roughly an
order of magnitude greater than that for SNIa
(Figure~\ref{fig:SNrates}).  The net result is that each type of
supernova is expected to produce roughly half of the observed $\Fe$.
With a parametrization for the supernova rates as a function of
redshift, and the mean amount of $\Ni$ produced for each type of
supernova, we can compare the expected supernova yields to the
observed $\Fe$ abundances.  Renzini~\cite{Renzini:2003} has compiled
measurements of $\Fe$ abundances in the intracluster medium of galaxy
groups and clusters, with an approximate average abundance of $30 \%$
solar by mass fraction.  Additionally, studies of the $\Fe$ abundance
in Damped Lyman-alpha systems show a large scatter in [Fe/H], with a
upper limit of roughly $10^{-3}$ as a fraction of the total baryonic
mass~\cite{Prochaska:2003hd}.  Given the difficulty in measuring the
$\Fe$ abundance, a comparison between the observed yields and that
expected to be contributed from supernovae is not very precise.
Nevertheless, assuming the current data for $\Fe$ cluster abundances
to represent the average abundance in the $z=0$
universe~\cite{Renzini:2003}, the observed $\Fe$ density parameter is
\begin{equation}
\Omega_{{\rm Fe, Obs}} = 0.3 \, Z_{\odot} \Omega_{b} \simeq 2 \times 10^{-5},
\end{equation}
with $Z_{\odot} = 0.0013$ the solar system $\Fe$ mass
fraction~\cite{Anders:1989zg}. For SNII, integrating the upper limit
curve in Figure~\ref{fig:SNrates} to obtain the total comoving number
density of SNII, and an average $\Ni$ production per event of $0.06 \,
{\rm M}_{\odot}$, the amount of $\Fe$ contributed by SNII is
$\Omega_{{\rm Fe, SNII}} \simeq 0.6 \times 10^{-5}$.  Similarly, from
the upper limit SNIa curve as constrained by the DSNB, and with $0.5
\, {\rm M}_{\odot}$ per SNIa event the $\Fe$ contribution is
$\Omega_{{\rm Fe, SNIa}} = 0.4 \times 10^{-5}$.  Thus in our
concordance model, the predicted $\Fe$ yield from SNII plus SNIa is
within a factor of two of the data, precluding the possibilities that
either supernova rate is significantly larger, or that both are
significantly smaller.  More precise measurements of $\Fe$ abundances
at all redshifts, when combined with tighter constraints on the $\Fe$
production per event, will strengthen the constraints on the supernova
$\Fe$ production.  Still, the existing crude constraint provides
additional evidence to support our concordance model.
  
\subsection{The Cosmic Far IR background} 
\label{sec:FIR}

While stars with mass greater than 8 $\Msun$ produce the DSNB flux,
the cosmic far infrared background flux is dominated by the absorption
of UV light on dust and re-emission at longer wavelengths from all
massive stars, making the observation of the FIRB flux limits an
independent determination of the CSFR in this stellar mass range. This
mass range is the same as that probed by SNIa, as well as by massive
x-ray binary systems~\cite{Norman:2004im}, and therefore the CSFR as
constrained by each of these proxies should be in agreement.  The
detailed analysis of the reprocessed starlight which accounts for the
FIRB will involve modeling the evolution of dust components and
temperature at all redshifts (e.g., ~\cite{Loeb:1997}).  The observed
FIRB intensity ($\gsim \, 100 \, \mu{\rm m}$) is ${\rm I}_{\nu} \sim
30\ {\rm nw}\ {\rm m}^{-2} {\rm sr}^{-1}$~\cite{Wright:2003tp}, a
result which is still subject to systematic errors from foreground
contaminations~\cite{Hauser:2001xs,Wright:2003tp}.  With reasonable
assumptions for the dust, the FIRB intensity is reproduced with a CSFR
similar to those measured by UV surveys, and so is consistent with the
concordance model presented here~\cite{Schiminovich:2004km}.  In
coming years the FIRB will be mapped by Spitzer from the near to far
IR, giving constraints on the CSFR as well as the population of IR
galaxies at high redshift~\cite{Spitzer}.

\section{The Cosmic Gamma-Ray Background}
\label{sec:CGB} 

In previous sections we have developed a concordance model of the CSFR
as constrained by the DSNB and light surveys, and from this predicted
the expected ranges for the evolution of supernova rates and other
indicators. Next we use our results to calculate the portion of the
MeV CGB made by SNIa. The connection between the DSNB and CGB has been
previously discussed by Hartmann et al.~\cite{Hartmann} and by Zhang
and Beacom~\cite{Zhang:2004tj}.  Additionally, Buonanno et
al.~\cite{Buonanno:2004tp} have recently explored the connections
between the neutrino and gravitational wave backgrounds, providing an
initial estimate of the expected gravitational wave background from
core-collapse supernovae.

\subsection{The CGB data} 

The CGB has been measured by the Solar Maximum Mission (SMM) in the
energy range $0.3-7$ MeV~\cite{Watanabe:1997} and the Compton imaging
telescope COMPTEL in the energy range $0.8-30$
MeV~\cite{Weidenspointner:1999,Kappadath:1998}.  The CGB analysis is
hindered by instrumental and cosmic-ray backgrounds, which must be
carefully subtracted to reveal the underlying signal.  Below 100 keV
and above 10 MeV, the diffuse background is expected to result from
the addition of individual unresolved point sources, with Seyfert
galaxies dominating at low energies and EGRET-detected Blazars
contributing at high energies.  Radio-loud Seyferts, such as Centarus
A, are observed with spectra that continue beyond 100 keV, though the
present density of such sources, combined with the extrapolation of
the spectrum tail to 10 MeV, do not imply a large contribution to the
diffuse background~\cite{G2001}. Additionally, the Blazar spectra
detected by EGRET, and confirmed by COMPTEL at lower energies, implies
a spectrum break and flux suppression below 10 MeV, likely ruling out
significant contribution to the MeV background~\cite{SS01} (however,
two AGN detected by EGRET and confirmed by COMPTEL are exceptionally
bright and time-varying sources at $1-10$ MeV~\cite{Blom}).
   
In the energy range $0.8-30$ MeV, the CGB spectrum fit by COMPTEL is
\cite{Kappadath:1998}
\begin{equation}
\frac{dN}{dt \, dA \, dE \, d\Omega}    =
(1.1  \pm  0.2)  \times  10^{-4}  \left(  \frac{E}{5  \,  {\rm  MeV}}
\right)^{-2.4 \pm 0.2} ({\rm MeV}^{-1} {\rm  cm}^{-2} {\rm  s}^{-1}  {\rm sr}^{-1}).
\label{eq:COMPTELfit}
\end{equation} 
This data is consistent, within large uncertainties, with a low energy
extrapolation from the EGRET background spectrum and high energy
extrapolation of the SMM spectrum, with an evolution from harder to
softer spectrum in the COMPTEL range.  It has been commonly accepted
that SNIa provide the primary contribution to the CGB in $0.8-3$ MeV
\cite{clayton91,Watanabe:1998ds,Ruiz-Lapuente:2000zx,Zhang:2004tj}.
In the remaining parts of this section we test this hypothesis.

%%%%%%%%%%%%%%%%%%%%%%%%%%%%%%%%%%%%%%%%%%%%%%%%%%%%%%%%%%%%%%%%%%
%%%%%%%%%%%%%%%%%%%%%%%%%%%%%%%%%%%%%%%%%%%%%%%%%%%%%%%%%%%%%%%%%%
\begin{figure}
\epsfxsize=4.5in
\begin{center}
\leavevmode
\epsffile{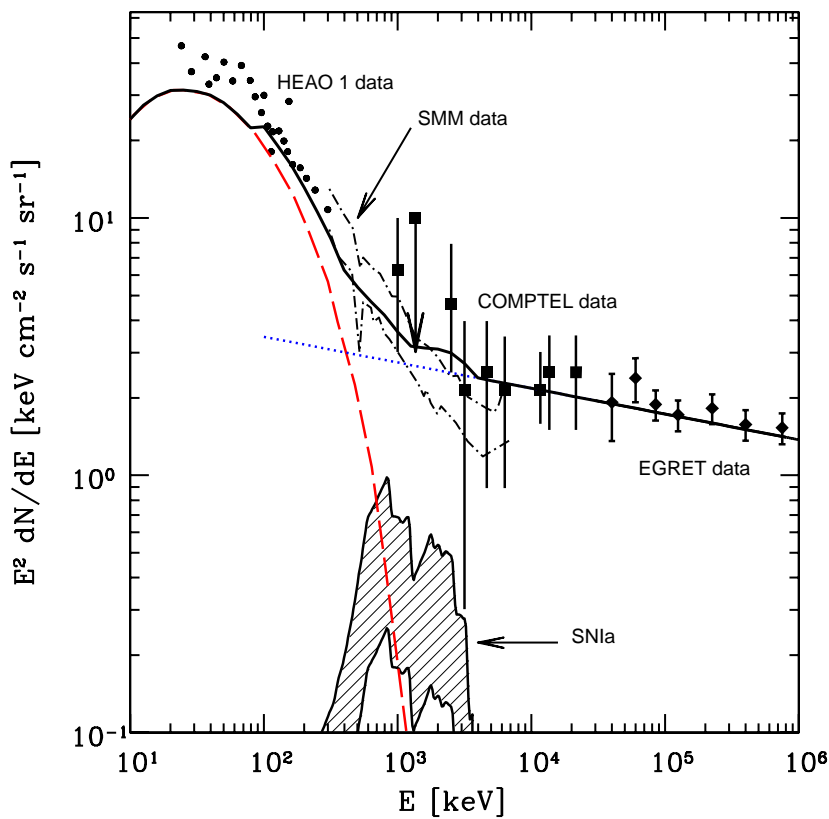} 
\end{center}
\caption{The observational results for the CGB 
  spectrum with the shading corresponding to the previous figures.
  The band shows the allowed contribution from SNIa, with the upper
  edge of the band preferred in our concordance model.  The data from
  COMPTEL are shown as squares~\cite{Weidenspointner:1999}, and SMM
  data range is given by the two dash-dotted
  lines~\cite{Watanabe:1997}.  Also shown are the HEAO x-ray data
  (circles)~\cite{Kinzer:1997}, and the EGRET data
  (diamonds)~\cite{Sreekumar:1998}.  The blue dotted line is the power
  law fit to the EGRET background and our assumed low energy
  extrapolation, and the red dashed curve is the expected contribution
  to the x-ray background from Seyfert galaxies~\cite{Zdziarski:1996}.
  The bold black curve is the sum of the Seyfert, SNIa, and Blazar
  spectra, with our assumed low energy extrapolation.  }
\label{fig:spectrum}
\end{figure}
%%%%%%%%%%%%%%%%%%%%%%%%%%%%%%%%%%%%%%%%%%%%%%%%%%%%%%%%%%%%%%%%%%%%%%%%
%%%%%%%%%%%%%%%%%%%%%%%%%%%%%%%%%%%%%%%%%%%%%%%%%%%%%%%%%%%%%%%%%%%%%

\subsection{The SNIa Gamma Ray Spectrum }
\label{sec:spectra}

During the thermonuclear explosion of a SNIa, $\Ni$ is produced as the
nuclear burning wave propagates through the white dwarf core. This
subsequently decays (half life $\sim 6$ days) to $\Co$ and then (half
life $\sim 77$ days) to stable $\Fe$; these decays are often to
nuclear excited states, and are thus accompanied by MeV gamma rays.
Most gamma rays from $\Ni$ are unable to escape from the dense
environment and are degraded into x-ray and lower energies. As the
optical light curve declines, the thinning outer layers are expected
to become optically transparent to gamma rays from the decay of $\Co$.
In total, SNIa are expected to produce $\sim 0.5 \, \Msun$ of $\Ni$,
with exact mass depending on the nature of the nucleosynthesis in the
core.

The SNIa gamma ray spectrum is thus characterized by lines which
encode the competition between the strength of the flux from $\Ni$ and
$\Co$ decays, and the opacity of the stellar material to gamma rays.
In our calculations we use the standard W7 model of Nomoto et
al.~\cite{Nomoto:1984sm}, as presented in Watanabe et
al.~\cite{Watanabe:1998ds}, for a representative spectrum. The
normalization of this spectrum is such that an average SNIa produces
$0.5 \, \Msun$ of $\Ni$, corresponding to $1.1 \times 10^{55}$ gamma
rays. Long-lived radioactive isotopes, such as $^{44}$Ti, $^{26}$Al,
$^{60}$Co, as well as positrons contribute to the gamma ray flux,
though at a level less by a few orders of magnitude.

Three nearby SNIa have been studied in gamma rays, resulting in upper
limits on the gamma ray flux $\sim 10^{-5} \, {\rm cm}^{-2} \, {\rm
  s}^{-1}$ for a SNIa at an assumed distance of 10
Mpc~\cite{Milne:2004mt}. These observed limits are very near
theoretical flux predictions, leading to constraints on the mass of
$\Ni$ produced in SNIa, and are consistent with our normalization of
the spectrum. These results indicate that the gamma ray emission from
SNIa is not {\it larger} than assumed here.

\subsection{The SNIa Contribution to the CGB}

The SNIa contribution to the CGB flux is determined directly from the
SNIa gamma ray spectrum and the SNIa rate.  The SNIa CGB flux is given
by Equation (\ref{eq:diffuseflux}), after accounting for the
convention that the CGB flux is typically quoted per solid angle.  The
shape of the SNIa CGB is determined primarily by the decay lines of
$\Ni$ and $\Co$. The spectrum cuts off above 3 MeV, since the
radioactive decay of these isotopes do not emit gamma rays above that
energy. At energies below 1 MeV, the spectrum blends into a hard x-ray
continuum below the threshold for gamma ray telescopes.  The spectral
features from the decay of $\Ni$ and $\Co$ are smoothed out by
integration over the SNIa redshifts, though these features could be
recovered by considering the angular correlations of the CGB
flux~\cite{Zhang:2004tj}.
         
In Figure~\ref{fig:spectrum} we show the SNIa CGB spectrum, with a
range determined by the uncertainty in the SNIa rate from
Figure~\ref{fig:SNrates}. The parameters for the SNIa rate have been
chosen as in Section~\ref{sec:SNIa}. We have determined a firm upper
limit to the normalization on the SNIa CGB from the concordance CSFR,
and this upper limit is nearly an order of magnitude below the CGB
data.  If SNIa had been the dominant source in the MeV range, spectral
breaks might have been seen, marking the transitions to other sources
at lower and higher energies. While the data can exclude strong
breaks, more mild transitions cannot be ruled out from the data alone.
Here we have used strong and independent evidence to exclude a
significant SNIa contribution to the MeV CGB. Still, a small and possibly detectable contribution
from SNIa may remain (see Figure~\ref{fig:spectrum}). 
         
\subsection{Comparison to Previous Results}  
 
The SNIa contribution to the CGB has been studied by previous authors.
Ruiz-Lapuente et al.~\cite{Ruiz-Lapuente:2000zx} compare the evolution
of the SNIa rate with CSFR models to conclude that reasonable models
exist in which the CGB can be fully accounted for by the SNIa
contribution.  The analysis of Watanabe et al.~\cite{Watanabe:1998ds}
concludes that SNIa may not be able to account for the entire CGB, but
uncertainties on ${\rm R}_{{\rm SF}} (z=0)$ at that time prevented
them from drawing stronger conclusions. Zhang and
Beacom~\cite{Zhang:2004tj} focused on angular correlations of the CGB,
and normalized their diffuse SNIa spectrum to match the COMPTEL data.

Given that these previous calculations use similar SNIa spectra based
on the models of Nomoto et al.~\cite{Nomoto:1984sm}, the primary
difference in these results is the adopted CSFR.  With a 1 Gyr delay
time, SCDM cosmology, peak redshift of star formation at $z=1$, and a
normalization of the SN rates such that $R_{{\rm SNIa}}/R_{{\rm SNII}}
= 1/3$, Watanabe et al. obtain a present SNIa rate of $R_{{\rm SNIa}}
(z=0) = 6 \, \times \, 10^{-5} \, {\rm yr}^{-1} \, {\rm Mpc}^{-3}$.
With 3 Gyr delay normalization to the SNII neutrino limit, and
efficiency determined from the evolution of the SNIa rate, we obtain
$R_{{\rm SNIa}} (z=0) = 3 \, \times \, 10^{-5} \, {\rm yr}^{-1} \,
{\rm Mpc}^{-3}$.  Ruiz-Lapuente et al. scan the space of allowed range
for the CSFR, but appear to have overestimated the SNIa gamma ray
spectrum by an order of magnitude (Figure 3
of~\cite{Ruiz-Lapuente:2000zx}), leading them to conclude that the MeV
CGB is entirely from SNIa. In this paper we show, even with the
maximum allowed CSFR, that the MeV CGB cannot be of SNIa origin.
  
\section{Discussion and Conclusions}
\label{sec:conclusions}
   
\subsection{New Sources for the Cosmic Gamma-Ray Background}  
  
Knowledge of the SNIa rates gives the best prediction yet for the SNIa
contribution to the MeV CGB.  With the SNIa rate which we have
determined, we can limit the observed SNIa contribution to the CGB to
nearly an order of magnitude less than observed.  Here we consider the
robustness of this estimate.  Our CSFR, as constrained by the DSNB,
reproduces the observed SNIa rate with reasonable choices of the SNIa
efficiency and time delay.  Our DSNB-constrained CSFR and SNIa
formation scenario is also in good agreement with a variety of
constraints on the light and Fe generated by stars of intermediate
mass and larger. In this sense, we have deemed this model
`concordance.'

In our analysis, the SNIa gamma ray luminosity assumed is the maximum
allowed by constraints from observations of individual SNIa
\cite{Dahlen:2004km}.  Keeping the gamma ray luminosities fixed, could
it be that SNIa are 10 times more abundant, either because they form
more efficiently and/or because the intermediate mass CSFR is larger
than we derive here?  Either of these possibilities would mean that
the SNIa are ten times more frequent than surveys indicate and that
the iron produced per SNIa must be several times less than assumed
here (in order to be consistent with the observed total iron
abundance), which is unlikely since the iron production per SNIa is
constrained by light curve energetics.  Given the IMF assumed above,
if the SNIa formation efficiencies are in the range used here, then
the CSFR would have to be an order of magnitude larger than we derive
from the DSNB.  The additional light generated from the increase in
intermediate mass and core-collapse mass stars would exceed the
measurements discussed here and, in addition, $90\%$ of massive stars
must be neutrino-impotent in order to not be observed by
Super-Kamiokande, an unlikely scenario (see Section \ref{sec:failed}).
These considerations lead us to conclude that SNIa do not
significantly contribute to the MeV CGB.

If the CGB is not due to SNIa, then what?  In Figure
\ref{fig:spectrum} we show three possible sources of to the CGB:
Seyferts, SNIa, and a somewhat arbitrary extrapolation of the $> 10$
MeV background (as measured by COMPTEL and EGRET)
\cite{Sreekumar:1997yg}. In combination, this three-component model
seems to be a reasonable fit to the CGB with the dominant contribution
at MeV energies arising from a newly postulated source: MeV Blazars,
by which we mean objects with emission spectra similar to the Blazars
thought to generate the $> 10$ MeV CGB, but now with numbers and
distribution sufficient to account for the observed MeV CGB intensity
\cite{Blom}.  We note that the Blazars required to generate the $> 10$
MeV CGB are expected to be cut off below $\sim$ 5 MeV and the sources
required to make the $\sim 1$ MeV CGB have yet to be observed. Such
MeV Blazars, or additional astrophysical sources, will be further
tested by future gamma ray telescopes (e.g., ACT, MEGA, and NCT
\cite{Milne02,Bloser:2001dj,NCT}).

Additionally, we can consider models for new physics which predict or
allow contributions to the present-day MeV CGB, for example the decay
of non-baryonic cold dark
matter~\cite{Kamionkowski94,Barbieri88,Gondolo92,Ellis92,Kribs97,Chen03,Abazajian01,Feng03,Olive85},
the decay of massive gravitons predicted by models of extra
dimensions~\cite{Arkani99,Hall99,Hannestad03,Casse03,Kolb03}, and
primordial black hole evaporation~\cite{Kim99,Kapusta02,Sendouda03}.
Photons from these sources may have a featureless energy spectrum
which is isotropic on the sky, and which could be distinguished from
candidate astrophysical sources by the angular correlations
\cite{Zhang:2004tj}. The uncertain origin of the MeV CGB opens up a
new window for these studies.

\subsection{Implications for Failed Supernovae}  
\label{sec:failed}
 
In our analysis, we have used observables and limits which probe the
star formation rate at different stages of stellar evolution. The UV,
optical, and IR observations probe the massive CSFR {\it before} the
neutrino emission from core-collapse, while the optical
measurements of the supernova rates and the heavy metal abundances
probe the CSFR {\it after} the neutrino emission. The optical
determinations of the supernova rates are particularly important. A
measurement of the ratio of the optical SNII to the SNIa rate in the
ranges determined in Figure \ref{fig:SNrates} implies that most SNII
are optically successful, with no significant failure rate due to
prompt black hole formation (after most of the neutrinos are emitted)
\cite{Heger:2002by,Beacom:2000ng,Beacom:2000qy}.  Good agreement
between these {\it before} and {\it after} observations, combined with
the results from SN 1987A, imply that there would be no astrophysical
mechanism which can suppress the DSNB flux. Novel neutrino properties,
e.g., invisible decays, could deplete the fluxes from astrophysical
neutrinos \cite{Beacom:2002cb,Beacom:2002vi,Beacom:2003nh}, and Ando
\cite{Ando:2003ie} and Fogli et al. \cite{Fogli:2004gy} have
considered the effects on the DSNB in detail.

This result is of high importance for the current detection of the
DSNB flux. With robust measurements of the optical SNII rate, the DSNB
flux could be predicted with no model dependence from the CSFR or IMF.
This should be an important priority for astronomical observations.
Given the few direct measurements for the optical SNII rate, however,
the DSNB flux is presently the most stringent limit on the
core-collapse and optical SNII rate.  With a core-collapse rate
derived from the CSFR, a measurement of the DSNB flux by itself can
determine the fraction of massive core-collapse events which fail to
emit neutrinos.  With the optical data as well, the DSNB flux would
also be the most robust measurement of the fraction of core-collapse
events which fail to explode as optical SNII.
 
\subsection{Detection of the Diffuse Supernova Neutrino Background}  
 
As we have discussed above, all indications from the CSFR observations
are that the DSNB flux {\it is} on the verge of detection, which has
been re-substantiated by the recent results from
GALEX~\cite{Schiminovich:2004km}, for which the best fit CSFR predicts
a nearly detectable DSNB.  Note that dust corrections larger than
indicated in Figure~\ref{fig:CSFRmodels} would begin to present a
serious inconsistency with the neutrino data. The current
configuration of Super-Kamiokande gives the best detection potential
for the flux through the $\bar{\nu}_e$ channel, though currently this
search is limited to the exponential tail of the spectrum at energies
greater than 19.3 MeV.  Loading Super-Kamiokande with gadolinium
trichloride would allow tagging of neutron captures, significantly
lowering backgrounds and the analysis threshold~\cite{Beacom:2003nk}.
This would allow a quick detection of the DSNB, and is the only
realistic possibility for measuring the DSNB spectrum
shape~\cite{Beacom:2003nk}.  Detection of the DSNB flux will be
important for studying the supernova rates, and could also be an
exciting milestone discovery of the first astrophysical source of
neutrinos beyond the Sun and SN 1987A.

\section{Acknowlegments}   

We thank Shin'ichiro Ando, Dieter Hartmann, Manoj Kaplinghat, and Mark
Leising for discussions. J.F.B. was supported by The Ohio State
University, and L.E.S. and T.P.W. were supported at OSU by the
Department of Energy grant DE-FG02-91ER40690. P.Z. was supported by
the DOE and the NASA grant NAG 5-10842 at Fermilab.

\end{document}